\documentstyle[12pt]{article}
\begin{document}
\topmargin 0cm
\baselineskip=.7cm
\parskip=.2cm
\textwidth=14cm
\textheight=21cm
\def\theequation{\arabic{section}.\arabic{equation}}
\newcommand{\beq}{\begin{equation}}
\newcommand{\eeq}{\end{equation}}
\newcommand{\rbracket}{\right]}
\newcommand{\fff}{{\bar f}}
\newcommand{\lbracket}{\left[}
\newcommand{\DD}{{\cal D}}
\newcommand{\DDt}{{\nabla}_{t}}
\newcommand{\ddt}{\partial_t}
\newcommand{\ddx}{\partial_x}
\newcommand{\h}{\theta}
\newcommand{\r}{\cal G}
\newcommand{\Ag}{A^{g}}
\newcommand{\aaa}{{\alpha}}
\newcommand{\Q}{\bar{Q}}
\newcommand{\Pb}{\bar{P}}
\newcommand{\ppb}{\bar p_{i}}
\newcommand{\ps}{\bar{\psi}}
\newcommand{\LL}{{\cal L}}
\newcommand{\ggg}{g}
\newcommand{\RR}{{\cal R}_{\nu_{i}}}
\newcommand{\ep}{\epsilon_{i}}
\newcommand{\Vq}{\Delta (\Q)}
\newcommand{\Vp}{\Delta (\Pb)}
\newcommand{\rw}{\rightarrow}
\def\op{operator}
\def\tly{topologically}
\def\mfti{Moscow Institute of Physics and Technology}
\def\onlabs{on leave of absence from}
\def\itep{ Institute of Theoretical and Experimental Physics}
\def\exl{external}
\def\CG{Clebsh-Gordon}
\def\gl{global}
\def\an{anomaly}
\def\f{field}
\def\exst{existence}
\def\nl{normalizable}
\def\nty{normalizability}
\def\cn{condition}
\def\con{configuration}
\date{
\setlength{\unitlength}{\baselineskip}
\begin{picture}(0,0)(0,0)
\put(9,20){\makebox(0,0){UUITP-6/93}}
\put(9,19){\makebox(0,0){ITEP-20/93}}
\put(9,18){\makebox(0,0){March 1993}}
\put(8,17){\makebox(0,0){hepth/yymmddd}}
\end{picture}}
\title{\bf HAMILTONIAN SYSTEMS OF  CALOGERO TYPE
           AND TWO DIMENSIONAL YANG-MILLS THEORY  }
\author{A. \ Gorsky
\thanks{\onlabs \itep, 117259,B.Cheremushkinskaya, Moscow,
Russia.}\\
and\\
N. \ Nekrasov
\thanks{\onlabs \itep, 117259,B.Cheremushkinskaya, Moscow,
Russia \newline and from the \mfti, Dolgoprudny, Moscow region, Russia}\\
           {\em \newline Institute of Theoretical Physics} \\
           {\em Thunbergsv\"agen 3 Box 803} \\
           {\em S-751 08 Uppsala Sweden}}
\maketitle
\begin{abstract}
We obtain integral representations for the wave functions of Calogero-type
systems, corresponding to the finite-dimensional Lie algebras, using exact
evaluation of path integral. We generalize these systems to the case of the
Kac-Moody algebras and observe the connection of them with the two
dimensional Yang-Mills theory. We point out that
Calogero-Moser model and the models of Calogero type like
Sutherland one can be obtained
either classically by some reduction from two dimensional
Yang-Mills theory with appropriate sources or even at quantum
level by taking some scaling limit.
We investigate large $k$ limit and observe a relation with
Generalized Kontsevich Model.

\end{abstract}
\section{Introduction}
\setcounter{equation}{0}
     It is well known that in some supersymmetric quantum mechanical theories
it is possible to evaluate path integrals exactly. Usually it is the case when
bosonic part of the action of this theory can be interpreted
as a Hamiltonian of circle action on the loop space, provided
that the symplectic structure on the loop space is defined by
the fermionic part of the action. Then algebra of the
supersymmetry is interpreted as an equivariant derivative
action and path integral defines an equivariant cohomology
class which in the nicest situation localizes on the fixed
points of circle action and gives finite-dimensional integral
as an answer. All this can be generalized to the case of non-abelian
group action \cite{atyah},\cite{witten}.

  It is a general belief that the localization technique can be applied
to the analysis of integrable systems. Namely it is expected that partition
functions and correlators in systems with finite or even infinite number
degrees of freedom can be calculated in this manner \cite{niemi}. So it is
natural to start from the quantum mechanical systems related with the finite
dimensional Lie groups. There are two large classes of such systems, namely
generalized Toda chains or lattices and generalized Calogero-Moser models.
Both of them posses a very transparent algebraic phase space and can be
described as a different projections of the free motion on the curved
manifold. These systems in some sense are the particular examples of the
multimatrix models but the underlying Hamiltonian structure of these models
provides one with a possibility to handle with them in a canonical manner.

   The multiparticle quantum mechanical systems evidently should be
generalized to the field theories. This program has been done for the
Toda system which smoothly transforms from quantum mechanics to the
Toda field theory. The underlying structure is the Kac-Moody
algebra, since the Poisson brackets in the field theory correspond
to the symplectic structure on the Kac-Moody coadjoint
orbit.
We shall show that there exists also the field theory generalization of
Calogero-Moser systems. It appears that these are nothing but
two-dimensional Yang-Mills theory with the appropriate sources.

The paper is organized as follows. In section $2$ we consider as an example
Calogero-Moser dynamical system and evaluate exactly the corresponding path
integral. Then we generalize this result to the case of generic root
system. In section $3$ we turn to the infinite-dimensional
situation and investigate two-dimensional Yang-Mills system and its
generalization in the  spirit  of  the  previous  constructions.  In
section $4$ we briefly discuss supersymmetric part of this stuff. In
section $5$ we discuss some open problems, concerning relations with
the rational solutions of KP hierarhies. In section $6$  we  present
our conclusions.

\section{ Calogero-Moser system }
\setcounter{equation}{0}

\subsection{Classical Calogero-Moser system as a hamiltonian reduction of the
free system}
Classical Calogero-Moser system is a system of N particles on the real line
with the pair-wise interaction potential \cite{moser}:
\beq \label{pot}
V(x_{1},\ldots,x_{N}) = \sum_{i\neq j}  \frac{g^{2}}{(x_{i}-x_{j})^{2}},
\eeq
$g^{2}$ is a coupling constant, which is supposed to satisfy $g^{2} > - 1/8$,
to avoid the collapse of the system.
It is well known \cite{per-ol},\cite{kag} that such a system (and all
generalizations, like Sutherland's one with $\frac{1}{sin^2 (x_{i}-x_{j})}$
instead of $\frac{1}{(x_{i}-x_{j})^{2}}$ or with extra quadratic potential)
appears as a result of hamiltonian reduction of some simple hamiltonian
system. For pure Calogero system (\ref{pot}) it goes as follows.
We start from the free system on the cotangent bundle to the Lie algebra
$su(N)$ . This means that we consider the space of pairs $(P,Q)$, where $P$
and $Q$ are $su(N)$ matrices (we identify Lie algebra $su(N)$ and the space of
hermitian matrices by multiplying by $\sqrt{-1}$) with
canonical symplectic structure
\beq \label{sym1}
\Omega = tr(\delta P \wedge \delta Q).
\eeq
Free motion is generated by quadratic hamiltonian
\beq \label{h2}
H_{2} = \frac{1}{2} tr P^2
\eeq
On the  $T^{*}su(N)$  acts unitary group $SU(N)$ by adjoint action on $Q$'s
and coadjoint on $P$'s. This action preserves  symplectic structure. If we
identify the cotangent bundle with two copies of the Lie algebra with help of
the Killing form we would get a moment map of this action in the form
\beq \label{mom1}
\mu = [P,Q]
\eeq
According to the standard prescription of hamiltonian reduction procedure we
restrict ourselves to some level $\mu$ of momentum map and then factorise
along the distribution of the kernels of the restriction of symplectic form.
Generally these ones are spanned by the vectors, tangent to the orbits of the
action of the stabilizer of point $\mu$ in the coadjoint representation.
In our situation stabilizer depends on the number of different eigenvalues of
matrix $\mu$.In particular, if matrix $\mu$ has only two different
eigenvalues, with multiplicities $1$ and $N-1$, then stabilizer will be
$S(U(1) \times U(N-1))$.
We shall denote it by $G_{\mu}$. Now we should resolve equation (\ref{mom1})
modulo transformations from this group. It is easy to show, that if matrices
$P$ and $Q$ satisfy condition $[P,Q]=\mu$ then we can diagonalize $Q$ by
conjugation by matrices from the $G_{\mu}$. Let $q_{i}$ be its eigenvalues.
Then $P$ turns out to be of the form (in the basis,where $Q$ is diagonal and
$\mu_{ij}=\nu(1-\delta_{ij})$)

\beq \label{ppp}
P=diag(p_{1},\ldots,p_{N})+ \nu ||\frac{1-\delta_{ij}}{(q_{i}-q_{j})}||
\eeq

where $\nu (N-1)$ and $-\nu$ are the eigenvalues of $\mu$.

Symplectic structure on the reduced manifold turns out to be the standard one,
i.e. $p$'s and $q$'s are canonically conjugate variables.
Hamiltonian (\ref{h2}) is now hamiltonian for the natural system
with potential $V$ (\ref{pot}). Note that if we replace momentum
matrix $\mu$ by the $F^{+} \mu F$ with $F$ being diagonal unitary
matrix, then  reduced Hamiltonian $H_{2}$ wouldn't change.
 In fact, it is possible to obtain integrable system, starting from the
Hamiltonian
\beq \label{hk}
H_{k} = \frac{1}{k} tr P^k
\eeq
All $H_{k}$ commute among themselves with respect to Poisson structure,
defined by (\ref{sym1}), thus yielding the complete integrability of the
classical Calogero model. The restriction to the $su(N)$ case imply just the
fixing of the center of mass of the system at the point zero.

\subsection{Path integral for the Calogero  model}

Now we proceed to the quantization of the Calogero system. We  consider path
integral representation for the wave functions and by means of exact
evaluation of path integral we  obtain finite-dimensional integral
formula for Calogero wavefunction (see \cite{per-ol-q}).
We shall consider integral over the following set of fields.
First, there will be the maps from the time interval $[0,T]$ to the direct
product of three copies of $su(N)$ Lie algebra, actually to the
product $T^{*}{su(N)}\times {su(N)}$. We denote corresponding
matrix-valued functions as $P(t)$,$Q(t)$ and $A(t)$
respectively. Then, we have fields $f(t)$ and $f^{+}(t)$ which are
the $CP^{N-1}$ - valued fields. Locally, these are $N$ complex numbers
$f_{i}$, which satisfy the condition $\sum_{i=1}^{N} f_{i}^{+} f_{i}
= 1$ and are considered up to the multiplication $f \rw e^{i \h} f$,
$f^{+} \rw f^{+} e^{-i \h}$.
 We don't explain right now the supersymmetric version of this
integral, so we just say that measure on $P$,$Q$ fields comes from the
symplectic structure (\ref{sym1}), the  $A$ field measure comes from the
Killing form on the Lie algebra, and the $f,f^{+}$ measure is again the
symplectic one, corresponding to the standard $CP^{N-1}$ Fubini -
Schtudi form $\Omega = p_{!} i^{*} \nu \frac{1}{2\pi i} \delta f^{+}
\wedge \delta f$.
Here, $i^{*}$ means the restriction on the sphere $S^{2N-1}: f^{+}f = N$, and
$p_{!}$ is the factorization along the $U(1)$ action $f \rw e^{i \h}
f$, $f^{+} \rw f^{+} e^{-i \h}$.
 Actually, as we will see, $A$ is not a scalar field, it is Lie algebra-valued
one-form (gauge field).
 More precise description of the measure will be given later,
in the section 4.

Now let us define the action. We shall do it in two steps. At first we define
the action for $P$, $Q$, $A$ fields. It will be a sum of action of
unreduced free system plus term which will fix the value of the
moment map. Field $A$ will play the role of the lagrangian
multiplier.

\beq \label{act1}
S_{P,Q,A} = \int tr ( i P{\ddt}Q - \frac{1}{2} P^2 + i A ([ P,Q ]  + i \mu))
\eeq

Since $\mu$ has only two different eigenvalues with multiplicities $1$ and
$N-1$, it has the following form:

\beq \label{mu}
\mu = i \nu ( Id - e \otimes e^{+} )
\eeq

where  $\nu$ is arbitrary (up to now) real number and we can choose vector $e$
to be $e = \sum_{i=1}^{N} e_{i}$, $e_{i}$ being the standard basis in $C^{N}$.

We shall evaluate the transition amplitude $<q'|\exp((t'-t)H)| q>$.
It is given by the path integral with boundary conditions.
First we integrate out the $A$ field. We observe that our path integral is
invariant under the (small) gauge transformations
\begin{eqnarray}
&&Q \rw g Q g^{-1} \; , \; P \rw g P g^{-1}
\nonumber\\
&&A {\rw} {\Ag} = g A g^{-1} + i({\ddt}g) g^{-1};
\nonumber \\
&&g(t) \in G_{\mu} \; , \; g(0) = g(T) = Id \label{gauge1}
\end {eqnarray}

We can enlarge the symmetry of our integral by introducing auxilliary fields
$f^{+},f$ as follows. Let us rewrite term $tr(\mu (\Ag))$ in terms
of the vector field ${\fff} = g^{-1} e$, with $e$ from the previous
chapter. We get
\beq \label{cpn}
S_{f} = \nu \int_{0}^{T} {\fff}^{+} (i \ddt - A ) {\fff}
\eeq

Obviously, $S_{f}$ depends only on the class of $\fff$'s  modulo
the total factor $e^{i \h}$ if at the ends $0$ and $T$ $\h$'s coincide.
Hence, it depends only on the $CP^{N-1}$ valued field $f$, which
corresponds to the $\fff$. Therefore,  our action $S = S_{P,Q,A} +
S_{f,f^{+}}$ accquires the form
\beq \label{actcpn}
S = tr ( i P {\DDt} Q - \frac{1}{2}P^{2} + i \nu f^{+} {\DDt} f ),
\eeq
where ${\DDt} Q = {\ddt} Q + i[ A, Q ]$, ${\DDt} f = {\ddt} f + i A f$.
It is clear, that this action has $SU(N)$ gauge symmetry:
\begin{eqnarray}
&& P \rw P^{g} = g P g^{-1}, \;  Q {\rw} Q^{g} = g Q g^{-1}
\nonumber\\
&&A {\rw} {\Ag} = g A g^{-1} + i({\ddt}g) g^{-1}
\nonumber \\
&&f \rw g f \; , f^{+} \rw f^{+} g^{-1} \; , g \in SU(N)
\label{gauge2}
\end {eqnarray}
This invariance holds in the sense that amplitude is invariant under the
$SU(N)$ action:
$<Q,f|\ldots|Q',f'> = <g Q g^{-1}, g f|\ldots| g' Q' g'^{-1}, g' f'>$.

As it usually goes in matrix models we diagonalize matrix $Q(t)$ by the
conjugation by unitary matrix $g(t)$. We denote diagonalized matrix $Q(t)$ by
$\Q$, and let $q_{i}$ be its entries.  Due to $SU(N)$ invariance of the action
"angle" variables $g(t)$ decouple except the boundary terms.
It is a gauge fixing so Faddeev-Popov determinant arises, which is nothing
but the square of the Vandermonde determinant of $q_{i}$ , ${\Vq}^{2} =
\prod_{i<j}(q_{i}-q_{j})^{2}$.

We can keep it just by  decomposing measure on the Lie algebra as the product
of the Cartan measure on the $q_{i}$ times ${\Vq}^{2}$ times the Haar measure
on the unitary group and it should be done in each point of time
interval, i.e. ${\DD}Q = \prod_{t} d{\Q} {\Delta({\Q}_{t})}^{2} dg$.
Actually we get here an extra integration over Cartan subgroup. This implies
that our gauge fixing is incomplete. We approve it by fixing the $f$ field
to be real and positive. This means that among all the representatives $ff$
of the point $f \in CP^{N-1}$ there should be one with real and positive
components. This can be always achieved in the generic point due to
the remainig action of the Cartan subgroup $T^{N-1}$.
It is interesting to note that factor-space $\Delta_{N} = CP^{N-1}/ T^{N-1}$
is nothing but the $(N-1)$ - dimensional simplex.
It is parametrized by the variables $x_{i} = f_{i}^{+}f_{i},\; i=1,\ldots,N$,
which satisfy conditions $x_{i} \geq 0 , \sum_{i=1}^{N} x_{i} = N$.
In fact, local coordinates on the $CP^{N-1}$ are provided by $x_{i}$ subjected
by just described conditions and ${\h}_{i}$ with conditions
$- \pi < {\h}_{i} <  \pi, \sum_{i} {\h}_{i} = 0$. These coordinates
are well-defined when point $x$ is inside of the internal region of
$\Delta_{N}$. The action $S_{f}$ can be rewritten as
$$
S_{f} = \int x_{i} {\ddt} {\h}_{i} - \sum_{i,j} \sqrt{x_{i} x_{j}}
e^{i({\h}_{i} - {\h}_{j})}A_{ij}
$$
Here, $A_{ij}$ are  matrix  elements  of  $A$  in  the  fundamental
representation. In our approach we fix the value of $x$ at the ends
of time interval. This choice of polarisation on $CP^{N-1}$ leads to
the wavefunctions with support on some special points $x$, such that
the monodromy of the $\nu$'th power of the Hopf line bundle
along the $N-1$-torus over $x$ is trivial (Bohr-Sommerfeld condition).
We will need the fact that after a quantization of $CP^{N-1}$ with
symplectic structure $\Omega$ some irreducible representation
$R_{\nu}$ of $SU(N)$ appears.
Its dimension
$$
dim R_{\nu} = \frac {(N + \nu -1)!}{\nu! (N-1)!}
$$
is the number of integer points in the corresponding simplex.
This information permits  us  partially  integrate  out  fields  in  our
theory, namely, we integrate $f^{+},f$ and this make us left with the
matix element of the ${\cal  P}  \exp  \int  A$  in  the  representation
$R_{\nu}$.
On the other hand, if we look at the integral over the non-diagonal part
of the $A$ field,  we  just  recover  constraint  (\ref{mom1})  and  the
Vandermonde - Faddeev - Popov determinant cancels.
Here we don't introduce ghost fields, but in principle it can be done and
the result will be the same.
We have gauged $Q$ to be in the Cartan subalgebra, therefore
the term $[P, Q]$ has become orthogonal to the Cartan part of the $A$.
This implies that integration out this part yields
$$
\prod_{t} \prod_{i=1}^{N-1} \delta( f^{+}_{i} f_{i} - 1) = \prod_{t}
\prod_{i=1}^{N-1} \delta( x_{i} - 1)
$$
This produces after integrating out $x_{i}$ fields some constraints
on the boundary values of the fields $x$. The symplectic measure on the
$CP^{N-1}$ factorised along the $T^{N-1}$ (i.e. the pushed forward
measure to the $\Delta_{N}$) is nothing, but
$\prod_{i=1}^{N} dx_{i} \times \delta(\sum_{i=1}^{N} x_{i} - N)$.
Geometric action $i\nu \int \delta^{-1} \Omega$ turns into
$i \nu \int \sum_{i} {\ddt} x_{i} = i \nu \sum_{i} x_{i}|_{t=T} -
x_{i}|{t=0}$.
So the integral over $CP^{N-1}$ valued part of the fields  is  done  and
the remaining integral over $P,Q,A$ recovers Calogero amplitude for
$g^{2} = \nu (\nu -1)$.

\subsection{Lattice version of the Calogero model path integral}

We wish to get the exact answer. To this end we present here
a lattice version of this path integral, which will be calculable and in
the reasonable continuum limit goes to the expressions written above.
The main motivation for this model will become clear below, in the section
concerning two-dimensional Yang-Mills theory. Let us point out, that this
construction resembles Migdal's calculation in \cite{migdal}.
So, in order to calculate a transition amplitude
$$
<Q',v'| \exp( - T H ) |Q'',v''>, Q \in su(N), v \in R_{\nu},
$$
we consider a time-like lattice $0 < t_{1} < \ldots < t_{K} < T$.
Let us denote its $i$'th vertex as $V_{i}$ and the link, passing
from the $V_{i}$ to $V_{i+1}$ will be denoted as $L_{i}$.
Actually we will calculate the answer for the
Sutherland's model with potential
$$
{\cal V} = \frac{(g/k)^{2}}{2 sin^2 \frac{(x_{i}-x_{j})}{k}},
$$
Thus we have introduced an extra parameter $k$ which will serve as
an interpolating parameter. Pure Calogero  model  arises  when  $  k  \rw
\infty$.
Our   variables   in   hands   will be the following:
elements   of   the   $SU(N)$
$g_{i}, h_{i}$, its  irreducible  representations  ${\aaa}_{i}$  and
vectors $|v_{i}>$ of some orthonormal basis in the  representation  $R_{\nu}$.
 $g_{i}$  and  $|v_{i}>$  are assigned at $V_{i}$, while ${\aaa}_{i}$
and $h_{i}$ live on $L_{i}$.
 Lattice version of the amplitude is given by the following formula:
\begin{eqnarray}
&&     <g_{0},v_{0}|      \ldots     |     g_{T},v_{T}>      =      \int
\prod_{L_{i},V_{i}}dg_{i} dh_{i} \sum_{\alpha_{i}} \sum_{v_{i}}
\chi_{\alpha_{i}} (h_{i} g_{i+1} \times
\nonumber\\
&&  \times  h_{i}^{-1}  g_{i}^{-1})   <v_{i}|T_{R_{\nu}}(h_{i})|v_{i+1}>
\times
\nonumber\\
&& \times dim(\alpha) \exp ( - \frac{c_{2}(\alpha_{i})}{k^{2}} \Delta t_{i})
\label{lattice}
\end{eqnarray}
It is supposed, that $g_{i} = \exp \frac{1}{k} X_{i} , X_{i} \in su(N)$,
$c_{2}(\alpha)$ is a quadratic  casimir  in  the  irreps  $\alpha$,  and
integration over $g,h$ goes with the Haar measure, normalised in such  a
way, that
$\int dg = 1$.
 If we would replace $c_{2}(\alpha)$ by any invariant polynomial on the
$su(N)$, we would get the answer for generalized Calogero model with
Hamiltonian $H = \sum_{k} {\epsilon}_{k} H_{k}$.
 Now let us consider the continuum limit of the (\ref{lattice}).
First (\ref{lattice}) is invariant with respect to the lattice
gauge transformations:
\begin{eqnarray}
&& g_{i} \rw U_{i} g_{i} U_{i}^{-1}, \; h_{i} \rw U_{i} h_{i} U_{i+1}^{-1};
\nonumber\\
&& |v_{i}> \rw T_{\RR}(U_{i}) |v_{i}>
\label{lattgauge}
\end{eqnarray}
 We can partially fix this gauge freedom (gaining group-like Vandermondes),
by diagonalizing $g_{i}$ and imposing some conditions on $|v_{i}>$.
Integration over remaining Cartan part of gauge transformations projects
vectors $|v>$ onto the subspace of the fixed points of Cartan torus action
on the $R_{\nu}$. For this subspace to be non-trivial, $\nu$ should be
divisible by $N$ \cite{alex2}.
 Let us consider contribution of one link. When $\Delta t_{i} \rw 0$ the sum
over representations turns into the $\delta ( h_{i} g_{i+1} h_{i}^{-1}
g_{i}^{-1} - Id)$. This implies that $g_{i}$ and $g_{i+1}$ have the same
eigenvalues. Generically, we could fix the gauge, ordering this eigenvalues.
So, in our gauge this yields $\delta(g_{i} - g_{i+1})$ and that $h_{i}$ is
in the Cartan subgroup, determined by $g_{i}$. In fact, we can set them to
be equal to $Id$ due to arguments from the previous paragraph. When
$\Delta t_{i} \neq 0$ but still remains to be small, we expect that the
main contribution to the integral (\ref{lattice}) comes from $g_{i}$'s which
slowly change, when $i$ varies, and from the $h_{i}$'s which are close to
unity, i.e. $h_{i} = Id - i \Delta t_{i} A_{i}$, $iA_{i} \in su(N)$.
Let us write $g_{i} = \exp ( i{\Q}_{i}/k)$. Then $h_{i} g_{i+1} h_{i}^{-1}
g_{i}^{-1}$ turns into
$$
\exp i({\Q}_{i+1} - {\Q}_{i})/k) - \frac{i\Delta t_{i}}{k} [A_{i},
e^{-i\frac{{\Q}_{i+1}}{k}}] e^{-i\frac{{\Q}_{i}}{k}}
$$
 Now we consider large $k$ limit (so we restrict ourselves to the Calogero
case. The treatment of the Sutherland system is simpler from the point of view
of Yang-Mills theory). We will use formula, which will be proven below,
in the chapter, concerning matrix models. Formula relates some limit
of the characters and the Itzykson-Zuber integral, namely
\beq \label{IZ}
lim_{k \rw \infty} \frac{\chi_{k \cdot {\aaa}}(\exp (i\frac{Q}{k}))}
{dim( k \cdot {\aaa})} = \int_{SU(N)} dU \exp (tr({\Pb}U{Q}U^{-1})
\eeq
where $k \cdot {\aaa}$ is the representation of $SU(N)$ whose signature
is $k \cdot p_{i}$ and ${\Pb} = diag (p_{1}, \ldots, p_{N})$.
In the limit $k \rw \infty$ sum over representations becomes the integral
over $\Pb$ with the measure $d{\Pb} {\Vp}^{2}$ (Vandermonde determinant comes
from the $dim(k \cdot)^{2}$. The integrand (contribution of
$V_{i},L_{i},V_{i+1}$) is
$$
\int dU \exp (tr (U^{-1}{\Pb}_{i}U ({\Q}_{i+1} - {\Q}_{i} + \Delta t_{i}
[A_{i}, {\Q}_{i+1}]) - \Delta t_{i} {\Pb}_{i}^{2} )) \times
$$
$$
\times <v_{i}|T_{R_{\nu}}({\cal P} \exp \int_{t_{i}}^{t_{i+1}} A)|v_{i+1}>
$$
 It is clear that after redefinition $P = U^{-1}{\Pb}_{i}U - \frac{1}{N}
tr({\Pb}) Id$ and rewriting matrix elements of ${\cal P} \exp$ in the
$R_{\nu}$ through $CP^{N-1}$ path integral, we just get the path integral
for the model (\ref{actcpn}) in the gauge $Q$ is diagonal.
 After we have convince ourselves, that in  the  limit  $\Delta  t_{i}  =
t_{i+1} - t_{i} \rw 0, k \rw \infty$ this sum goes to the desired path
integral with  the action (\ref{actcpn}), we can integrate out field $g_{i}$.
It turns out that integration over $g_{i}$, assigned to some vertex $V_{i}$,
gives us after the redefinition $h_{i} \rw h_{i-1}h_{i}$ the expression of
the  same  type,  as (\ref{lattice}), so we can remove all internal
vertices and we are left with the following integral representation
for the amplitude:
\begin{eqnarray}
&&|{\Delta}(\exp(i{\Q}/k))| \sum_{\aaa}
dim(\aaa) \int dh \chi_{\aaa}(g_{0} h g_{T}^{-1} h^{-1})
\nonumber\\
&& \times \exp( - \frac{T}{k^{2}} c_{2}(\aaa)) <v'|T_{R_{\nu}}(h)|v''>
\label{amplitude}
\end{eqnarray}
 This expression can be calculated even in the case  of  finite  $k$  and
answer  involves  {\CG}  coefficients. It is obvious that if $|v'>$ and
$|v''>$ are not the fixed vectors of the Cartan torus action on $R_{\nu}$
then the amplitude vanishes.
 States are enumerated by the representations $\alpha$ ( and by basis vectors
in $Inv({\aaa} \otimes {\aaa}^{*} \otimes R)$). The signature of the $\alpha$
divided by $k$ gives the asymptotic momenta of the particles.
One more observation concerning this simple example is that our wavefunction
deals with a zonal spherical functions.
In the zero coupling limits $\nu \rw 0,1 $ wavefunctions turns into
$$
\Psi_{\nu = 0} =\frac{1}{N!} \sum_{\sigma \in {\cal S}_{N}} \exp (i\sum_{l}
\Pb_{\sigma (l)} \Q_{l})   $$
 and
$$
\Psi_{\nu = 1} = \frac{det|| \exp (i\Pb_{l} \Q_{m})||}{{\Vp}}
$$
 respectively (${\cal S}_{N}$ is a Weyl group for $SU(N)$) \cite{per-ol-q}.At
generic coupling constants the system manifests the anyonic behaviour
\cite{alex3}

   Let us mention that the integration over $Q$ would lead to
the Lax description of the integrable system. Lax equation implies that the
Hamiltonian flow goes along the coadjoint orbit with the Hamiltonian $A$.
It is clear that it is not the only way of the integration in this functional
integral. For example, $P$ field can be integrated out first.
For the  Hamiltonian $H_{2}$ we have a gaussian integral, and one immediately
gets the particular example of the $c=1$ gauged matrix model as an answer.
Now the motion is not restricted on the one orbit and keeping in mind
relation between the coadjoint orbits and representations of the group
we can speak about the flow between the representations (actually, between
all, not necessarily unitary ones). Thus the motion goes in the model space.

\subsection{Calogero systems, related to generic root system}

 In this section we shall generalize this result to the case of generic root
system. Calogero system, considered above, corresponds to the root system
$A_{N-1}$.
 In the next paragraph we remind the notion of the root system
\cite{heckman},\cite{arnold} and, following \cite{per-ol}, describe
corresponding integrable systems.

 Let $(V,<,>)$ be the vector space over $R$ with inner product and let
$\Delta$ be a finite set of vectors (roots) satisfying the following
conditions:

\beq \label{rootdef}
\mbox{if} \; \alpha,\beta \in \Delta \; \mbox{then} \;
r_{\aaa}(\beta):= \beta - 2 \frac{<\aaa,\beta>}{<\aaa,\aaa>}\aaa \in
\Delta ;
\eeq
\beq
\mbox{and}  \frac{2<\aaa,\beta>}{<\aaa,\beta>} \in {\bf Z}
\eeq

Operators $r_{\aaa}$ entering into the definition are refered to as
reflection in the root ${\aaa}$, hyperplane, orthogonal to the root
${\aaa}$ is called the mirror, corresponding to the ${\aaa}$. Root system
${\Delta}$ is refered to as simple root system if it cannot be decomposed into
the direct sum of the two orthogonal root systems.

There is a classification of simple root systems, for example, the $A_{N-1}$
root system is a set of vectors $e_{ij}= e_{i} - e_{j}, 1 \leq i,j \leq N$ in
the $R^{N}\cap (\sum_{i} x_{i} = 0 )$.
With any semi-simple Lie algebra some root system is related (see
\cite{heckman}) and with any root system we can associate a dynamical system
on $V$ with the Hamiltonian
\beq \label{rootham}
H_{\Delta} = \sum_{i=1}^{N} \frac{1}{2} p_{i}^{2} + \sum_{\alpha \in \Delta}
\frac{g_{\aaa}^{2}}{<q,\aaa>^{2}}
\eeq
Here $q=(q_{1},\ldots,q_{N}) \in V$, $g_{\aaa}^{2} = \nu_{\aaa} (
\nu_{\aaa} - 1) $ is a coupling constant and for the roots, belonging to the
same orbits of the Weyl group, the value of the coupling constant should be
the same. Actually, here also generalizations, including $sin(<q,\alpha>)$ or
hyperbolic functions and/or extra quadratic potential are possible (in what
follows we will call these generalizations as corresponding to the curved
case).

Each such a system can be obtained as a result of hamiltonian reduction of the
free system on the cotangent bundle to the corresponding Lie algebra. Extra
Hamiltonians $H_{k}$ (\ref{hk}) are replaced by the generators of the ring of
invariant polynomia on the Lie algebra. Corresponding value of the momentum
map is chosen as follows:

\beq \label{rootmom}
\mu = \sum_{\aaa} \nu_{\aaa} (e_{\aaa} + e_{-\aaa})
\eeq
Here $e_{\aaa}$ are the roots of the corresponding Lie algebra.

Quantum calculation proceeds as before, the only difference is that instead
of the group $SU(N)$ we have now generic semi-simple Lie group.

\section{Two-dimensional Yang-Mills theory as a field theory limit of the
Calogero system}
\setcounter{equation}{0}

In this section we will describe two things: first, we describe natural large
$N$ limit of the Calogero system, by that we mean replacing $SU(\infty)$ group
by the central extension of the loop group ${{\LL}SU(N)}$ -
$SU(N)$ Kac-Moody (KM) algebra. We shall see that such a theory describes
two-dimensional
Yang-Mills theory with external source on a cylinder. Then we shall show, that
compactification of the cylinder yields back $N$-particle Calogero system.
  The general procedure of deriving the field system starting from $SU(N)$
Calogero system should go as follows. At first one should consider the
$SU(\infty)$ root system which is the two dimensional lattice on the complex
plane (it corresponds to the representation of the $SU(\infty)$ as the
${\LL}SU(\infty)$).We can conjecture that the corresponding Calogero system has
Weierstrass $\wp$
function  as a potential. Then the $SU(N)$ KM root system embeds into lattice
as the finite lenght strip of the roots. It is important that the KM spectral
parameter which serve as a parameter on the loop appears already in the
Baker-Akhiezer function for the $\wp(x)$ potential \cite{krichever}. In what
follows we shall omit these intermediate steps and start with the KM algebra
from the very begining, postponing the establishing of the relation between
the $\wp$-function - type potential Calogero models and that one we shall
obtain here.

\subsection{Two-dimensional Yang-Mills theory via hamiltonian
reduction}

In the preceding chapters we have described construction of hamiltonian
system, correspondimg to some semi-simple Lie algebras ${\ggg}$ (in
fact, only invariant non-degenerate inner product was needed). We
considered "free" system on the cotangent bundle $T^{*}{\ggg}$ with
Hamiltonian $H = V(P)$, where $P \in {\ggg}^{*}$ (on the same footing
we could consider Hamiltonian, which is a function on the ${\ggg}$
instead of ${\ggg}^{*}$) and $V$ is some invariant polynomial on $\ggg$. Then
we applied the procedure of hamiltonian reduction, corresponding
to the adjoint - co-adjoint action of Lie group $G$ on the  $T^{*}{\ggg}$.
In order to quantize this reduced system we have introduced $T^{*}{\ggg}$ -
valued fields and some auxilliary fields $A(t) \in {\ggg}$ and
$f(t), f^{+}(t) \in X_{J}$, with $X_{J}$ denoting the co-adjoint
orbit of the value $J \in {\ggg}^{*}$ of
moment map over which we wished to make the reduction.
Here we generalize this construction to the infinite-dimensional case, namely
we would like to substitute finite-dimensional Lie algebra by
the Kac-Moody one.

 We start from consideration of the central extension of
current algebra ${\LL}{\ggg}$, where $\ggg$ is some semi-simple Lie algebra.
The cotangent bundle to this algebra consists of the sets $(A,k;\phi,c)$,
where $\phi$ denotes $\ggg$-valued scalar field on the circle, $c$ is
a central element, $A$ is $\ggg$-valued one-form on the circle and $k$
is the level, dual to $c$.
The symplectic structure is the straightforward generalization of
(\ref{sym1}):
\beq \label{sym2}
\Omega = \int tr (\delta A \wedge \delta \phi) + \delta k \wedge \delta c
\eeq
Corresponding action of the loop group has the following form \cite{samson}:
an element $(g(x),m| x \in S^{1})$ acts as follows:
\beq \label{km-adj}
\phi \rw g \phi g^{-1} , \; c \rw c + \frac{1}{2\pi} \int dx tr \phi g^{-1}
{\ddx} g
\eeq
it was an adjont part of the action, and the co-adjoint one is
\beq \label{km-coadj}
A \rw g A g^{-1} - \frac{k}{2\pi} {\ddx}g \cdot g^{-1}, \; k \rw k
\eeq
This action preserves the symplectic structure (\ref{sym2}) and the
corresponding moment map
\beq \label{momym}
(J,{\bar k}) = (\frac{k}{2\pi} {\ddx} \phi + [ A, \phi], 0)
\eeq
has vanishing level.
This implies that the infinite-dimensional analogue of $X_{J}$ is highly
degenerate one, and co-adjont orbits of such a type have functional
moduli space, namely the space of maps from the circle $S^{1}$ to
the Cartan subalgebra of $\ggg$ modulo the global action of
$G$ Weyl group, ( the  symmetric  group  ${\cal  S}_{N}$  for  $G  =
SU(N)$).

Now let us write down path integral for this system. To this end we need some
extra fields: $A_{0}$ as a Lagrangian multiplier, and some fields which
describe coordinates on the co-adjoint orbit of some element
$J \in {\ggg}^{*} \otimes \Omega^{1}(S^{1})$. Let us denote this orbit
as $X_{J}$, and corresponding field will be $f$. Let $\Omega_{J}$ be
the symplectic structure on $X_{J}$ and let $\mu_{J}$ be the moment
map, corresponding to the co-adjoint action of ${\LL}G$ on $X_{J}$.
Then our action is
$$
S = S_{\phi,A} + S_{J}
$$
where
\beq \label{actym1}
S_{\phi,A} = \int \int_{S^{1} \times [0,T]}dxdt tr ( i \phi \ddt A_{1} -
\frac{\epsilon}{2} \phi^{2}
+ A_{0} ( \frac{k}{2\pi} {\ddx} \phi + [A_{1}, \phi ])
\eeq
and
\beq \label{actym2}
S_{J} = \int \int_{S^{1} \times [0,T]}(\delta^{-1} \Omega_{J} - dt A_{0}
\mu_{J})
\eeq
Action $S$ is clearly invariant under the action of ${\LL}G$, provided that
$\Omega_{J}$ is integral, i.e. the orbit $(X_{J}, \Omega_{J})$ is quantizable.
Let us parametrize the orbit $X_{J}$ by the group element $g(x,t)$.
We perform a change of variables $A_{1} \rw \frac{1}{k} A_{1}$,
it gives us the following action $S$:
\beq \label{ym}
S = \int_{\Sigma} tr(i k \phi F - d \mu \frac{\epsilon}{2} \phi^{2}
- J(x) g^{-1} ({\ddt} + A_{0}) g)
\eeq
Here $\Sigma$ is some two-dimensional surface with distinguished measure
$d\mu$
and the choice of the coordinates $x,t$ can be considered as some choice of
polarization.
For example, it is possible to choose the holomorphic one, i.e. substitute
$(x,t)$ by $(z, {\bar z})$. In this case $J$ has to be holomorphic $\ggg$
valued one-form.  $J$ can be always gauged to be the Cartan element
(  see above the description of degenerate orbits ).
Now we have to define the measure. To be consistent with
finite-dimensional example we should define it in a way which distinguishes
$A_{0}$ and $A_{1}$. We prefer to follow \cite{witten} and define the
measure for fields $A_{i}, i=0,1$ and $\phi$ as follows.
The space of all $G$ connections on the $\Sigma$ is a symplectic one, with
symplectic structure
\beq \label{sym3}
\Omega = \frac{k}{4\pi^{2}}\int_{\Sigma} tr (\delta A \wedge \delta A)
\eeq
It implies that on $A$'s there exists a natural symplectic measure.
The measure on $\phi$ comes from the norm:
\beq \label{met}
||\delta \phi||^{2} = \int_{\Sigma} d\mu \; \frac{1}{2} tr \;\delta \phi^{2},
\eeq
which depends only on the measure $d \mu$ on the surface $\Sigma$.
A question about the measure on $f(x,t)$ could be solved in the same way, i.e.
we could choose a symplectic measure on the orbit. The problem is
that the naive measure is just the $\prod_{x,t} dg_{J}(x,t)$ where
$dg_{j}$ is the (finite-dimensional) measure on the co-adjoint orbit
$X_{J(x)}$ of the $J(x)$ in $\ggg^{*}$.
In general, dimension of $X_{J(x)}$ could be changed when one pass
from one point $x'$ to another $x''$.
Another problem is that possible quantum correction could destroy
the invariance of the theory. So we prefer to choose here an
appropriate ansatz for $J(x)$. Below we consider the case $G = SU(N)$,
${\ggg} = su(N)$.
 First, we remind that Calogero-type systems correspond to the $J$ with
maximal proper stabilizer. Let us take

\beq \label{ansatz}
J(x)dx = \nu(x)dx (N \cdot Id - e(x) \otimes e^{+}(x)),
\eeq

where ${\nu}(x)dx$ is some one-form on the circle and $e(x)$ is some field
in the fundamental representation of $SU(N)$, which can be gauged as $e(x) =
\sum_{i=1}^{N} e_{i}$, where $e_{i}$ denotes a standard basis in $C^{N}$.
The stabilizer of this element is the loop group ${\LL}S(U(1) \times
SU(N-1))$.
This choice doesn't resolve the problem of the measure. We restrict our
consideration to $\nu = \sum_{i=1}^{l} \nu_{i} \delta (x - x_{i})$.
Now we have an orbit $X_{J} = \times_{i} (CP^{N-1}, \Omega_{\nu_{i}})$,
where $\Omega_{\nu_{i}}$ is the standard symplectic form on
$CP^{N-1}$ $\omega_{F}$,
multiplied by $\nu_{i}$. Let $\RR$ be the representation, which corresponds '
to $(CP^{N-1}, \Omega_{\nu_{i}})$ after quantization by Kirrilov construction.
Now we can take measure on $X_{J}$ to be
$\prod_{i} \Omega_{\nu_{i}}^{\wedge N-1}$.
This will give us a desired path integral measure.
States in this theory will contain vectors $\otimes_{i} |v_{i}>$ in the
tensor products of representations $\RR$. In fact, states are
functionals $\Psi(A)$ of the gauge field $A_{1}$ on the circle, multiplied
by the vectors $|v_{i}>$, invariant under the action of gauge group:
\begin{eqnarray}
&&\Psi(A) \otimes \otimes_{i} |v_{i}> \rw
\nonumber\\
&& \rw \Psi(\Ag) \otimes \otimes_{i} T_{\RR} (g(x_{i}) |v_{i}>
\label{gauge3}
\end{eqnarray}
Here $T_{\RR}(g)$ is an image of element $g$ in the representation $\RR$.
It is obvious, that the only gauge invariant of the gauge field
$A_{1}$ in the interval
$(x_{i}, x_{i+1})$ is the monodromy
$$
h_{i,i+1} = {\cal P} \exp \int_{x_{i}}^{x_{i+1}} A_{1}
$$
It is invariant of the gauge transformations, which are trivial at the ends of
interval $(x_{i},x_{i+1})$. So, wavefunctions are functions of $h_{i,i+1}$
and we must take invariants of the $SU(N)$ action
$$
\Psi(h_{i,i+1}) \otimes \otimes_{i}|v_{i}> \rw
$$
$$
\rw \Psi(g_{i} h_{i,i+1} g_{i+1}^{-1})
\otimes \otimes_{i}T_{\RR}(g_{i})|v_{i}>
$$
For example, if we have only one representation $R$, i.e. $l = 1$,
then the Hilbert space is just
$$
\oplus_{\aaa} Inv( {\aaa} \otimes {\aaa}^{*} \otimes R)
$$
where sum goes over all irreps of $SU(N)$.
When $R$ is trivial, this gives us just the space of all characters.

Now we proceed to the calculation of transition amplitudes.
It is easy to show that it is given by the path integral over the
fields $\phi, A$ on the cylinder $S^{1} \times (0,T)$ with fixed
boundary monodromies $h_{i,i+1}$ and vectors $|v_{i}>$:
\begin{eqnarray}
&&< \{ h_{i,i+1}', v_{i}'\}|exp(-TH)|\{ h_{i,i+1}'', v_{i}''\} > =
\nonumber\\
&&\int {\DD}A {\DD} \phi \exp \int tr (ik \phi F -
\frac{\epsilon}{2} \phi^{2}) \times
\nonumber\\
&&\times \prod_{i} <v_{i}'| {\cal P} \exp \int_{C_{i}} A |v_{i}''>_{\RR},
\label{gym-path}
\end{eqnarray}
here $C_{i}$ is a straigth contour on the cylinder, which goes along
the time axis from the point $(x_{i},0)$ to the point $(x_{i},T)$.

This path integral can be easily reduced to the finite-dimensional one,
using techiniques, developed in \cite{witten}, \cite{migdal}, \cite{gross}.
This reduction appears due to the possibility of cutting the cylinder along
the contours $C_{i}$. Then, path integral on the disk $D_{i,i+1}$
with boundary $\partial D_{i,i+1} = C_{i+1} - ([x_{i},x_{i+1}], T) - C_{i} +
([x_{i},x_{i+1}],0)$,
(signs + and - mean orientation) is given by the following sum
\begin{eqnarray}
&&I_{D_{i}} = \sum_{\aaa} \chi_{\aaa}(h_{i,i+1}' g_{i+1} h_{i,i+1}''^{-1}
g_{i}^{-1})
\nonumber\\
&& \times e^{-\frac{TL_{i,i+1}}{k^{2}} c_{2}(\aaa)}
\label{plaquette}
\end{eqnarray}
where $L_{i,i+1} = x_{i+1} - x_{i}$, $c_{2}(\alpha)$ is the quadratic casimir
in the irreps $\aaa$, $g_{i}$ is the monodromy along the $C_{i}$ and
over $g_{i}$ we must integrate with the Haar measure. So the answer has
the following form:
\beq
\int \prod_{i} dg_{i} <v_{i}'|T_{\RR}(g_{i})|v_{i}''> I_{D_{i}}
\eeq
In principle, this integral over product of group manifold can be done and
the final answer includes the Clebsh-Gordan coefficients. The
general formula is rather complicated.
The partition function calculation will involve integrals of the
products of characters of representations $\RR$ and $\alpha$, hence
the answer will include multiplicities of the enterings of representations
$\RR$'s into the tensor products ${\aaa} \otimes {\aaa}^{*}$.
Let us denote them as $D({\RR},{\aaa})$.
We observe that the spectrum of this theory depends on the $\bf Q$ - relations
between the $L_{i,i+1}$'s and $D({\RR},\alpha)$. If the latters are $\bf Q$ -
independent, then for generic $R$ the spectrum is given by
$$
E_{\alpha_{1}, \ldots, \alpha_{l}} = \sum_{i} \frac{L_{i,i+1}}{k^{2}}
c_{2}(\alpha_{i})
$$
If lenghts $L_{i,i+1}$ are linearly independent over $\bf Q$, then
the spectrum is quasi-continious (for this effect to take place $l$ have to
be greater then one).

In the case $l = 1$, the case of the only one representation $R$,
we have an integral of the type
$$
I(h',h'')_{\alpha, R} = \int dg \chi_{\alpha}( h' g h'' g^{-1})
<v'|T_{R}(g)|v''>
$$
It is given by the sum over the orthonormal basis $\h$ in the space of
all invariants $Inv(\alpha \otimes \alpha^{*} \otimes R)$:
\begin{eqnarray}
&&\sum_{\h} C_{<i_{\alpha}| \otimes <j_{\alpha^{*}}| \otimes <v'_{R}|}^{\h}
<j_{\alpha}|h'|i_{\alpha}> \times
\nonumber\\
&&\times
C_{<k_{\alpha^{*}}| \otimes <m_{\alpha}| \otimes <v'' {R^{*}}|}^{\h}
<k_{\alpha}|h''|m_{\alpha}>,
\label{l1case}
\end{eqnarray}
where $C_{\ldots}$ are in fact the {\CG} coefficients and $|i_{\aaa}>,
 \ldots, |m_{\alpha}>$ run over basis in the irreps $\alpha$, while
 $|i_{\alpha^{*}}>$ etc. run over the dual basis.

It explains the correspondence between the Calogero-Moser model,
the Sutherland model and Yang-Mills theory with inserted Wilson line.
To get the Calogero answer we
must take $h',h''$ to be diagonal matrices
$$
h = diag ( \exp (i \frac{q_{i}}{k}))
$$

and take large $k$ limit, keeping ${\Pb}_{i} = \frac{\ppb}{k}$ finite,
where $\ppb$'s are numbers, connected with the lenghts of the columns of
the Young tableau of $\alpha$ (see below). Then we will recover
wavefunction of the Calogero model, corresponding to the state with
asymptotic momenta ${\Pb}_{i}$.

 Let us justify once  again  this  correspondence  by  resolving  the
constraint, coming from the fixing the value of the  moment  map  at
the point $J(x) = \nu \delta(x) (Id - e \otimes e^{+})$ ($l = 1$). We can
gauge field $A_{1}$ to the constant diagonal matrix, and from
constraint after simple calculations we conclude, that diagonal part
of $\phi$ is some constant map from the circle to the Cartan
subalgebra - $\phi_{ii} = p_{i}$, while non-diagonal part is  given
by the following formula:
\beq \label{ndiagphi}
\phi_{ij}  =  \nu  \lbracket  sgn(x)  -   cot(\frac{\pi   (q_{i}   -
q_{j})}{k}) \rbracket e^{-\frac{2\pi i x}{k} (q_{i} - q_{j})}
\eeq
where $q_{i}$ are eigenvalues of $A_{1}$,, $0< x \leq 1$ is  a
coordinate  on  the  circle   and   $sgn((x)$   is   a   multivalued
step-function. This expression is periodic, though it jumps when $x$
pass through some point $x = 0$ on the circle. This gives a
Hamiltonian
\beq \label{suther}
H_{2}   =   \sum_{i}\frac{p_{i}^{2}}{2}   +    \sum_{i    \neq    j}
\frac{(\nu/k)^{2}}{2 sin^{2}(\frac{q_{i}-q_{j}}{k})}
\eeq
In fact, more rigorous proof involves the consideration of
two Wilson lines in the same representation. Then constraint can be
resolved in the class of continious $\phi$'s. In the limit when the
distance between lines goes to zero we obtain just {\ref{suther}}. This
establishes  the  desired  relation  once  again.  (As  usual,
reduction at classical level doesn't impose any  conditions  on  the
$\nu$ and we don't see the quantum shift $\nu^{2} \rw \nu (\nu-1)$.)

\subsection{Relation with matrix models}

Let us investigate in more details the dependence on the $k$ parameter.
In what follows we set all sources to be zero, so we  take  pure  YM
system.
We see that if we make a change of variables
$$
A_{1} \rw \frac{1}{k} A_{1}, \; \psi_{1} \rw \frac{1}{k} \psi_{1}
$$
 then path integral accquires the following form:

\beq \label{path3}
k^{-N}\int {\DD}A{\DD}\psi{\DD}\phi \exp ( \int_{\Sigma}  tr (
k(\frac{1}{2}\psi \wedge \psi + i \phi F )) + \frac{\epsilon}{2} \phi^{2}
)
\eeq

Here $\psi$ is a fermionic field, which is a superpartner of A, i.e.
odd-valued one-form in adjoint representation. It serves here just for the
proper definition of the measure ${\DD}A$.
 After simple integrating out $\phi$ field integral turns into
two-dimensional QCD path integral:

\beq
k^{-N}\int {\DD} A \exp ( \int_{\Sigma} tr (- \frac{k^{2}}{2 \epsilon} F_{\mu
\nu}^{2})))
\eeq

The thing which deserves some comment is the appearence of the factor
 $k^{-N}$ (it is on the genus one surface, in case of genus $g$ the
factor will be $k^{-(N^{2}-1) (g-1)}$). Actually it comes from the
simple observation that if one considers an integral over symplectic
manifold ${\cal M}$ of the form $\exp (\omega)$ then $\int \exp (k
\omega) = k^{dim{\cal M}/2} \int \exp(\omega)$. In our situation
integral partially localizes on the moduli space of flat connections
in the principal $SU(N)$-bundle over $\Sigma$. We might
say that we have normalized our path integral in such a way that its
perturbative in $\epsilon$ part, which is responsible only for the integral
over the moduli space remains finite then $k \rw \infty$. It is easy to show
that if we take our space-time $\Sigma$ to be a torus of area $\int_{\Sigma}
d\mu = T$ then in the absence of the source $J$ the partition function is
given by the simple generalization of the Migdal-Witten formula
\cite{witten},\cite{migdal},\cite{rusakov} :

\beq \label{part}
Z = k^{-N}\sum_{\alpha} \exp ( - \frac{\epsilon T}{k^{2}}c_{2}(\alpha))
\eeq

where sum goes over all irreducible representations of $SU(N)$ and
$c_{2}(\alpha)$ is a quadratic casimir in the representation $\alpha$. The
wave function in the $q$ representation is now a function of the monodromy $U
= g_{A}(2\pi)$, invariant under conjugation, so it is a character and traces
in the irreducible representations provide basis in the Hilbert space. Each
such a trace is a common eigenfunction for all invariant polynomia on the
$su(N)$ Lie algebra considered as an operators in the theory.
This statement is general and is valid in any two-dimensional gauge theory. We
can consider slightly generalized Yang-Mills system (GYM), which
corresponds to the inclusion of the higher hamiltonians (\ref{hk}).
This system has a form

\beq \label{ggym}
{\cal L} = \int_{\Sigma} tr ( k \phi F + \sum_{i=2}^{N} {\ep}
\frac{\phi^{i}}{i} )
\eeq

Let us consider the solution of the corresponding Schr\"odinger equation with
initial condition $\Psi(U)|_{t=0} = \delta ( U - Id )$, $U \in SU(N)$. It is
easy to show, using \cite{witten}, that

\beq \label{psi}
\Psi(U)|_{t=a} = \sum_{\alpha} \exp ( -a \sum_{i} \frac{\ep}{i k^{i}}
c_{i}(\alpha)) dim(\alpha) \chi_{\alpha} (U).
\eeq

Now we would like to consider large $k$ limit. For this we shall make a small
digression and remind some simple facts about the irreps of the $SU(N)$
(\cite{gel}).
Each finite-dimensional irreducible representation $\alpha$ of the $SU(N)$
group is characterised by the set of integers $(p_{1} \geq , \ldots , \geq
p_{N} \geq 0)$ - the signature of the representation. Let us introduce
conventional variables ${\ppb} = p_{i} + N - i$. Then dimension, character
and casimirs of the representation $\alpha$ are equal to

\begin{eqnarray}
&&dim ( \alpha) = \frac {\Delta (\ppb)}{\Delta (i)};
\nonumber\\
&&\chi_{\alpha}(U) = \frac{det||{\beta_{j}}^{\ppb}||}{\Delta(\beta_{j})}, \;
\mbox{where}\; U = g\; diag(\beta_{1},\ldots,\beta_{N}) g^{-1},\; g \in SU(N);
\nonumber\\
&&c_{l}(\alpha) = \sum_{i=1}^{N} {\ppb}^{l} \prod_{j \neq i} ( 1 -
\frac{1}{\ppb -{\bar p_{j}}} )
\label{irreps}
\end{eqnarray}

Generic connection on the circle can be gauged to the diagonal
constant matrix $A = diag( iQ_{1}, \ldots, iQ_{N})$, therefore the
value of the wavefunction at the generic point
$U = diag (\exp( \frac{2\pi i}{k} Q_{j}))$ is

\beq \label{psik}
\Psi (Q_{j},a) = \frac{k^{\frac{N(N-1)}{2}}}{\Delta( Q_{i})} (1 + {\cal
O}(\frac{1}{k})) \sum_{\ppb} \exp ( -a \sum_{l} {\ep} \frac{P_{l}^{i}}{i})
\frac{det|| \exp ( i P_{i} Q_{j})||}{\Delta(P_{j})}
\eeq

where $P_{i} = \frac{\ppb}{k}$. In the limit $k \rw \infty$ this sum turns
into the integral over the ${\Pb}$, which is  nothing  (modulo  some
trivial factors, coming from  the  slightly  different  integration
region), but the partition function for the Generalized Kontsevich
Model (GKM) (see \cite{morozov1},\cite{morozov2}). Here we have used
Itzykson-Zuber integration formula
$$
\int_{SU(N)} dU \exp(tr(U{\Pb}U^{-1}{\Q})) = \frac{det||e^{iP_{i}Q_{j}}||}
{{\Vp}{\Vq}}
$$

It would be interesting to realize, what kind of integrable  system,
what kind of $\tau$ - function arises if Wilson loops are included.
Let us mention (it follows easily from the \cite{witten}) that
perturbative in $\ep$ part of the partition function for GYM is a generating
function for intersection pairings of cohomology classes of the moduli space
of flat $SU(N)$ - connections, namely

\beq \label{pair}
Z_{pert} = k^{-N}\int_{\cal M} \exp ( k \omega + \sum_{i} \ep \Theta_{i} )
\eeq

where $\omega$ is the symplectic structure of $\cal M$ and $\Theta_{i}$ are
characteristic classes of the tautological bundle ${{\cal A}_{0}} \rw {\cal
M}$, with ${\cal A}_{0}$ being the space of all flat $SU(N)$-connections on
the $\Sigma$.

It is interesting to note that $k \rw \infty$ limit yields the
quasi-continious spectrum of $\frac{c_{l}(\alpha)}{l k^{l}}$ and gives back
the spectrum in the Calogero model, corresponding to the $tr ( \frac{1}{l}
{\Pb}^{l})$. Additionaly, we see the interpolation between the wave function
in the Calogero model with zero coupling $\Psi_{\Pb}$ and the wave function in
the 2D YM theory  $\chi_{\alpha}$ where $\alpha$ is a representation, which
appears after quantization of the $SU(N)$ co-adjoint orbit of $\frac{\Pb}{k}$.

Let us again consider two dimensional Yang-Mills theory on the cylinder
$\Sigma = S^{1} \times (0,T)$. We normalize measure $d\mu$ in such a way that
the whole area of $\Sigma$ is just $2\pi T$. We have seen in the previous
subsection that in the limit $k \rw \infty$ wavefunctions of the YM system go
into the quantum mechanical wavefunctions, corresponding to the system of $N$
free particles on the real line, so we have got a kind of compactification in
2d YM theory.

In principle, it is not a trivial question whether it is possible to define
the procedure of compactification for topological theory, like topological
Yang-Mills theory or even for the physical two dimensional gauge theory which
is invariant under the action of the area preserving diffeomorphismes. In that
kind of theory, small radius of the spatial direction corresponds to the small
area of the space-time manifold - the only (except genus and number of holes)
invariant of two dimensional symplectomorphismes group. In 2d YM theory
small area limit corresponds to the contibution of the moduli space of flat
connections to the partition function.

  Let us look at the duality picture between the motion along and across the
orbits in the KM case. It is easily seen that integrating over ${A_{1}}$ we
get the motion along the $\phi$ (adjoint!) orbit. On the other hand if we
integrate over $\phi$ from the very beginning then usual YM-like (and higher
order) terms in the curvature appear. $k \neq 0$ coadjoint KM orbits
are labelled by the conjugacy classes of the finite dimensional group,
so if it happens that there is a flow of these classes then we have got the
motion, transversal to the orbits. Here it is just the case because we have
nontrivial in general evolution of the eigenvalues of the monodromy matrix.

\section{Supersymmetry}
\setcounter{equation}{0}

It is well known that the safest way to deal with path integral measure is to
rewrite it using superpartners of matter fields. For example, 2d YM partition
function is rewritten as \cite{witten}:

\beq
Z = \int {\DD} A {\DD} \psi {\DD} \phi \exp ( -\int tr (\frac{1}{2}\psi
\wedge
\psi + i \phi F + \frac{\epsilon}{2} \phi^{2}))
\eeq

 In the case of the YM system with the source $J$ which is not conserved in a
sense of the $SU(N)$ gauge theory, the procedure must be more elaborated.
In what follows we consider finite-dimensional case, i.e. the Calogero system.
The natural supersymmetrisation of quantum mechanical system involves
introducing of the superpartners of the variables $P,Q$ in such a way that
zero modes of them represent one-forms on the symplectic manifold. Keeping
in mind the equivariant $SU(N)$ derivative, acting
on the equivariant forms on the $T^{*}su(N) \times CP^{N-1}$ and extending it
to the loop space, we get the following action of the supersymmetry algebra:

\begin{eqnarray}
&& {\cal Q} P = \psi_{P}, \;  {\cal Q} Q = \psi_{Q}, \; {\cal Q} A = 0
\nonumber\\
&& {\cal Q}\psi_{P}= i {\DDt} P, \; {\cal Q}\psi_{Q} = i {\DDt} Q - P,
\nonumber\\
&& {\cal Q} f = \Psi_{f}, \; {\cal Q} f^{+} = \Psi_{f}^{+}
\nonumber\\
&& {\cal Q} \Psi_{f} = i {\ddt} f , \; {\cal Q} \Psi_{f}^{+} = i {\DDt} f^{+}
\label{supersymf1}
\end {eqnarray}

Path integral measure
$$
\DD \mu = {\DD} P {\DD} Q {\DD} A {\DD} \psi_{P} {\DD} \psi_{Q}
{\DD} f^{+} {\DD} f {\DD} \Psi_{f}^{+} {\DD} \Psi_{f}
$$
is well-defined due to supersymmetry.

Action, which includes fermions, can be written (locally) as
$$
S = \{ {\cal Q} , V \},
$$
with
\begin{eqnarray}
&& V = \frac{1}{2} \int ( tr(P \psi_{Q} - Q \psi_{P}) +
\nonumber\\
&& + \Theta_{i} \Psi_{f_{i}} - \Theta_{\bar i} \Psi_{f_{i}}^{+} )
\end{eqnarray}
It gives
\begin{eqnarray}
&& S = \int (tr(\psi_{P} \psi_{Q}) + \Omega_{{\bar i}j} \Psi_{f_{i}}^{+}
\Psi_{f_{j}}
\nonumber\\
&& + i tr(P{\DDt}Q - \frac{1}{2}P^{2}) + i f^{+}{\DDt}f\\
\label{superaction}
\end{eqnarray}
where $\Omega_{{\bar i}j}$ is a symplectic form on the $CP^{N-1}$, i.e.
Fubini-Shtudi form,
multiplied by $\nu$, and $\Theta$ is a (locally defined)
symplectic potential, i.e.
$$
\Omega_{{\bar i}j} = {\partial_{\bar i}} \Theta_{j} -  {\partial_{j}}
\Theta_{\bar i}
$$.
In general, our operator $\cal Q$ is not nilpotent, but it is nilpotent when
restricted onto the space of observables, invariant under the action
of some group, which is defined below.
This proves the equivariant closeness of the action and permits to use the
localization techniques to evaluate path integral (at least,
partition function).

Due to general arguments, we can add to action any term $t \{ \cal Q , V' \}$
and the answer will be independent of $t$, provided that we calculate
expectations of $\cal Q$ closed objects. This gives the way of exact
evaluation of path integral \cite{tirkkonen}.
The equivariant derivative which we have written corresponds
to the action of the semidirect product of $\cal R$ and ${\cal L} SU(N)$
groups on the loop space ${\cal L} T^{*}su(N) \times CP^{N-1}$.
An element $(g(t), \tau), g(t) \in {\cal L} SU(N), \tau \in {\cal R}$ acts
as follows:
$$
Q(t) \rw g(t) (Q(t + \tau) - \tau P(t + \tau)) g(t)^{-1},
P(t) \rw g(t) P(t + \tau) g(t)^{-1}
$$
$$
f(t) \rw g(t) f(t + \tau), f^{+}(t) \rw f^{+}(t + \tau) g(t)^{-1}
$$
Loop space (pre)symplectic forms $\int tr (\psi_{P} \psi_{Q})$ and $\int
\Omega_{{\bar i}j} \Psi_{f_{i}}^{+} \Psi_{f_{j}}$
are invariant under this action.
They can be equivariantly extended to become equivariantly closed forms.
This extension is nothing, but our action.

It is quite surprising from the finite-dimensional point of view
that in the case of the reduction under the zero level of the moment map
there is another realisation of the supersymmetry algebra,
which corresponds to the dimensional reduction of the YM theory.
We know that in this situation we don't need extra degrees of freedom,
corresponding
to the co-adjoint orbit of the moment value, since it vanishes. So we
are left with the fields $P,Q$ and $A$. The action reads as follows:

\beq \label{supercal}
Z = \int {\DD} P {\DD} Q {\DD} A {\DD} \psi_{Q} {\DD} \psi_{A} \exp ( \int tr
( i P {\ddt} Q + \psi_{Q} \psi_{A} + A [P,Q]  - \frac{\epsilon}{2} P^{2}))
\eeq
$\epsilon$ is arbitrary constant.

Second supersymmetry algebra is realized as follows:

\begin{eqnarray}
&& {\cal Q} A = \psi_{A},  {\cal Q} Q = \psi_{Q}, {\cal Q} P = 0
\nonumber\\
&& {\cal Q}\psi_{A}= i {\DDt} P, {\cal Q}\psi_{Q} = [ Q, P ],
\label{supersymf2}
\end {eqnarray}

\section{Relation with the rational KP and KdV solutions}
\setcounter{equation}{0}

It is known for a long time \cite{krichever} that the rational, trigonometric
and elliptic solutions of KdV and KP are closely related with the Calogero
system, for example the rational KdV solution is

\beq
u(x,t)=\sum_{i} \frac{g_{i}^{2}}{(x-x_{i}(t))^{2}}
\eeq
where the motion of the poles goes according to the Calogero Hamiltonians
\cite{per-ol}.

More generally if one has some elliptic spectral curve  $\r$ with the periods
related to the periods of the corresponding Weierstrass function and maps it
into the covering algebraic curve ${\r}_{n}$ defined by the  characteristic
equation

\beq \label{cur}
det({\lambda}  + L(x_{i}(t),{\aaa}))=0
\eeq

where $L$ is Lax operator, (the matrix $P$ in our notations ),

\beq \label {fir}
L_{ij} = p_{i} {\delta}_{ij}+2(1-{\delta}_{ij}) F(x_{i}-x_{j},{\aaa})
\eeq
\beq \label{sec}
F(x,\aaa) = \frac{{\sigma(x-{\aaa})} {e^{\xi(\aaa) x}}} {{\sigma(\aaa)}
{\sigma(x)}}
\eeq
\beq \label {th}
 {\xi}(z) = \frac{{\sigma}^{\prime}(z)}{{\sigma}(z)}
\eeq

$\aaa$ is  the point on the spectral curve, $\sigma$ is the Weierstrass
function, then this curve can be mapped into $n$-dimentional torus whose
coordinates are nothing but the angle variables for the $n$-particle system
with pairwise interaction via the Weierstrass potential. The zeroes of the
elliptic KdV tau function which is the $\h$ function of the curve $G_{n}$
now define the coordinates of the particles

\beq \label{zero}
\prod_{i} {\sigma}(x-x_{i}(t)) = 0
\eeq

The same is true for the KP elliptic solutions. The number of the particles
equals to the genus of the curve ${\r}_{n}$. The degeneration of the spectral
curve when one or both periods tends to infinity gives rise to the
trigonometric or rational interaction potentials. The transition from the
trigonometric to rational case can be interpreted as the transition from the
group to the algebra \cite{alex1}. We don't have at a moment the transparent
picture for the transition from the elliptic to trigonometric case. From our
consideration it follows that $k$ dependence could, in principle, play the
role of the interpolating parameter between $\wp$, inverse sine squared  and
inverse square potential, or, between Lie algebra and Lie group,
since the relevant group elements in hands look like
$diag(e^{\frac{2\pi i}{k} Q_{i}})$, which is a group-like element
for finite $k$ and seems to be Lie algebraic one in the limit $k \rw \infty$

  Another point which needs further clarification is the connection of the
KdV or KP structures with the two-dimentional YM. In fact the evolution
of the rational KdV solution is the motion along the Virasoro
coadjoint orbit which is the dual object to the space of the complex
structures of Riemann surfaces and the corresponding GKM models feel
the structure of this moduli space. From
the other hand two dimentional YM feels the structure of the moduli
space of the flat connections which was mentioned above. Therefore
having the relation between Calogero models and rational or in
general elliptic solution of KdV   from the one hand and with YM
from other hand we should find the relation
between these two objects. It should be formulated in terms of the relation
between the moduli space of the flat connections and the moduli space of
Riemann surfaces. The motion corresponding to the rational solutions have to
be reformulated as some motion in the phase space of YM.

\section{Conclusions}

Let us  summarize the picture just considered. We have shown that the
proper generalization of the Calogero-Moser system leads to the two
dimentional YM theory with a source. There is a relation between the wave
functions of Calogero and compactified YM.  So having this relation at hands it
is natural to ask all
the standard YM questions in terms of Calogero-Moser and vise versa.
  Another point is the path integral definition for the (at least finite
dimensional) integrable systems.  Different choises of the polarization of the
phase space, i.e. different integration orders in the functional integral,
manifest the duality structure of the systems at hand - we saw directly two
different flows along and across the orbits.

Many important questions remained beyond the scope of the paper.It is
important to realize the proper meaning of the generalization of
Calogero-Moser to the curved case,the corresponding picture for the Virasoro
algebra,introduction of the oscillator potention which has the meaning of the
mass term  in YM and so on.We will consider it in futher publications.The
problems related with the anyonic interpretation and the relation with the
collective field theory \cite{avan} have not been discussed.

   A point which also to be mentioned is that in the infinite coupling constant
limit in
the Calogero system with the oscillator term the particles are localized at
the equilibrium points and thus we have no nontrivial dynamics in the space
of the observables. It is known that the equilibrium position for the system
above and the system of the Coulomb particles in the oscillator potential
coincide so in the infinite coupling constant limit the systems are closely
related. From the other hand the Coulomb particles  represent the equivalent
picture for correlators in two dimentional CFT so the Calogero particles
imitate the positions of the vertex operators. Thus the infinite coupling
constant limit corresponds the case of the fixed positions of vertex operators
on the world sheet or in other words we have no moving branching points.

When the paper was completed,we have become noticed about the works
\cite{minahan},\cite{douglas} on related subjects.

\section{Acknowledegments}

We are grateful to V.Fock, A.Gerasimov, S.Kharchev,D.Lebedev, A.Lossev,
A.Marshakov,
A.Morozov, A.Mironov, A.Niemi, M.A.Olshanetsky, A.Rosly, G.Semenoff,
S.Shatashvili and A.Zabrodin for stimulating and helpful discussions. We
acknowledge Institute of Theoretical Physics at Uppsala University and Prof.
A.Niemi for kind hospitality. One of us (N.N.) is grateful to Proff.
H.Duistermaat, V.Guillemin, G.Heckmann and P.van Moerbeke for the nice
lectures at the Fall school on geometry of hamiltonian systems at Utrecht,
Nov.,1992 where some ideas of this work have appeared.

\end{document}